\RequirePackage{fix-cm}
\documentclass[twocolumn]{svjour3}          
\smartqed  
\usepackage{graphicx}
\usepackage{braket}
\usepackage{amsfonts}
\usepackage{cite}
\usepackage{amsmath}

\begin{document}

\title{Quantum Support Vector Machines for Continuum Suppression in B Meson Decays
}
\subtitle{}


\author{Jamie Heredge         \and
        Charles Hill      \and
        Lloyd Hollenberg    \and
        Martin Sevior       
}


\institute{Jamie Heredge, Charles Hill, Lloyd Hollenberg, Martin Sevior \at
              School of Physics, University of Melbourne, Parkville, VIC, 3020, Australia \\
              \email{jamie.heredge@student.unimelb.edu.au}             \\
              \email{cdhill@unimelb.edu.au} \\
              \email{lloydch@unimelb.edu.au}  \\
              \email{martines@unimelb.edu.au} 
           \and
           Charles Hill \at
                School of Mathematics and Statistics, University of Melbourne, Parkville, VIC, 3010, Australia \\
                \email{cdhill@unimelb.edu.au} 
}

\date{Received: date / Accepted: date}

\maketitle

\begin{abstract}
Quantum computers have the potential to speed up certain computational tasks. A possibility this opens up within the field of machine learning is the use of quantum techniques that may be inefficient to simulate classically but could provide superior performance in some tasks. Machine learning algorithms are ubiquitous in particle physics and as advances are made in quantum machine learning technology there may be a similar adoption of these quantum techniques.
In this work a quantum support vector machine (QSVM) is implemented for signal-background classification. We investigate the effect of different quantum encoding circuits, the process that transforms classical data into a quantum state, on the final classification performance. We show an encoding approach that achieves an average Area Under Receiver Operating Characteristic Curve (AUC) of 0.848 determined using quantum circuit simulations. For this same dataset the best classical method tested, a classical Support Vector Machine (SVM) using the Radial Basis Function (RBF) Kernel achieved an AUC of 0.793. Using a reduced version of the dataset we then ran the algorithm on the IBM Quantum {\textit{ibmq\_casablanca}} device achieving an average AUC of 0.703. As further improvements to the error rates and availability of quantum computers materialise, they could form a new approach for data analysis in high energy physics.

\end{abstract}

\section{Introduction}
\label{intro}
There are a number of measurements in flavour physics that are statistically limited due to lack of precision in signal-background classification. A superior classification algorithm would allow improved measurements and may result in the discovery of physics beyond the standard model. With the emergence of programmable quantum computer devices, we design and implement a quantum support vector machine approach for the signal-background classification task in B meson decays. In a B meson factory an electron and positron are collided, in our dataset this results in the creation of either a pair of B and anti-B mesons ($B\bar{B}$) decayed from the $\Upsilon(4S)$, or a lighter quark and anti-quark pair ($q\bar{q}$). The $B\bar{B}$ pair event is referred to as a signal event; the $q\bar{q}$ pair event is referred to as a continuum background event. The task is to classify events as either signal or background.

The $B\bar{B}$ will quickly decay and the $q\bar{q}$ will hadronise into an array of other longer-lived particles which are measured by the Belle II detector. To investigate a specific B meson decay mode, such as $B \to K^+ K^-$, we select only events that contain particle tracks identified as $K^+ K^-$ that could have originated from B mesons (which will include some falsely identified $K^+ K^-$ tracks from the $q\bar{q}$ pair), we refer to these particles as from the B candidate. If we use particles from the B candidate itself to perform the classification then we run the risk of sculpting the background to look like signal \cite{HAWTHORNEGONZALVEZ201954}. We therefore exclude particles associated with the B candidate and use the variables from the other B meson which are not correlated with the kinematic variables of the signal B. This limits the possibilities of sculpting our background to look like our signal. Therefore, given the momentum data of all the other final particles in the event (excluding those from the B candidate), we need to classify between the two scenarios of an initial $B\bar{B}$ signal pair or $q\bar{q}$ background pair. The $q\bar{q}$ continuum background events comprise the primary background for many studies of B-meson decays, many of which have branching ratios smaller than $10^{-5}$. It is therefore important to suppress the presence of continuum background events in the data. A better classification algorithm between signal and background events enables improved precision in measurements of these rare B-meson decays. 

Quantum computer hardware is advancing rapidly. Quantum supremacy has been achieved \cite{Google_suprecemecy2019, wu2021strong} by demonstrating a calculation on a quantum machine that outperformed classical high performance computers. Additionally, a photonic quantum system achieved a sampling rate of order $10^{14}$ above state-of-the-art simulations, completing a task that would be estimated to take current supercomputers several billions of years \cite{Hefei}. 
As the size and quality of quantum computers continues to advance, with large-scale entanglement achieved on a range of platforms \cite{mooney2021generation}, so does the feasibility of using quantum machines to perform classification tasks in particle physics. In particular, there are various ways in which the field of machine learning may benefit from the advent of quantum computing. These benefits range from speed-ups to specific subroutines, such as gradient descent \cite{rebentrost2018quantum}, to quantum analogues of classical algorithms, for example quantum neural networks \cite{quantum_neural_review}. We focus on the quantum analogue of the support vector machine that is proposed by Havlicek et al \cite{havlicek2018supervised}, specifically the kernel estimation technique. This QSVM approach involves using a quantum circuit to estimate the inner product between two datapoints that have been encoded into a higher dimensional quantum Hilbert space. This forms a kernel matrix which is then passed to a classical support vector machine. QSVM approaches have been found to outperform classical SVMs on various machine learning benchmark datasets \cite{park2020practical} and techniques involving preprocessing for QSVM approaches have led to improved performance on Character Recognition datasets \cite{yang2019support}. 

It has been suggested that variational quantum machine learning models can be fundamentally formulated as quantum kernel methods \cite{schuld2021quantum}. The global minimum of the cost function for a given quantum model is therefore defined by the kernel and thus the data encoding strategy. This highlights the importance of the data encoding step in any quantum model. Techniques such as quantum metric learning have demonstrated trainable data encoding strategies that aim to maximise distance between separate classes in the higher dimensional Hilbert space \cite{lloyd2020quantum}. In this paper we explore various data encoding methods that aim to capture the underlying structure of the data while also being easy to implement on current quantum machines.

Machine learning is widely employed in High Energy Physics \cite{albertsson2019machine}. Classically the signal-background classification task has been tackled with the use of constructed variables. This involves creating metrics from the momentum data, such as Fox-Wolfram moments, which are then used as inputs to a classical machine learning algorithm e.g. a boosted decision tree \cite{keck2016fastbdt, PhysRevLett.91.261801}. Our focus is on using the raw momentum data as the input to a quantum algorithm. This quantum algorithm then generates a kernel matrix that can be passed to a classical algorithm (a Support Vector Machine) to perform the classification. This approach allows us to test whether quantum circuits are capable of generating useful novel encodings for this classification problem.  

It has been shown that quantum machine learning techniques can be used for the discrimination of interesting events from background \cite{Terashi_2021, wu2020application, Wu_2021}. Alternative applications of quantum algorithms within particle physics have also included particle track reconstruction, utilising both quantum annealers \cite{bapst2019pattern} and quantum neural networks \cite{belayneh2019calorimetry}. A review of quantum machine learning in particle physics was carried out by Guan et al \cite{Guan_2020}.

\section{Problem Statement}
\label{sec:problem_statement}
Excluding the particles from the B candidate, our final dataset consists of $p$ particles with known momentum and therefore $3p$ inputs corresponding to the $3$ momentum components of each particle. Our aim is to produce an algorithm that takes particle momenta as an input and outputs a classification score of 0 for background events and 1 for signal events. The two performance metrics we report are the accuracy (percentage of correct classifications from all predictions) and the Area Under Receiver Operating Characteristic Curve (AUC).

\section{Implementation}
\label{sec:implementation}
\begin{figure}
\includegraphics[width=\linewidth,clip]{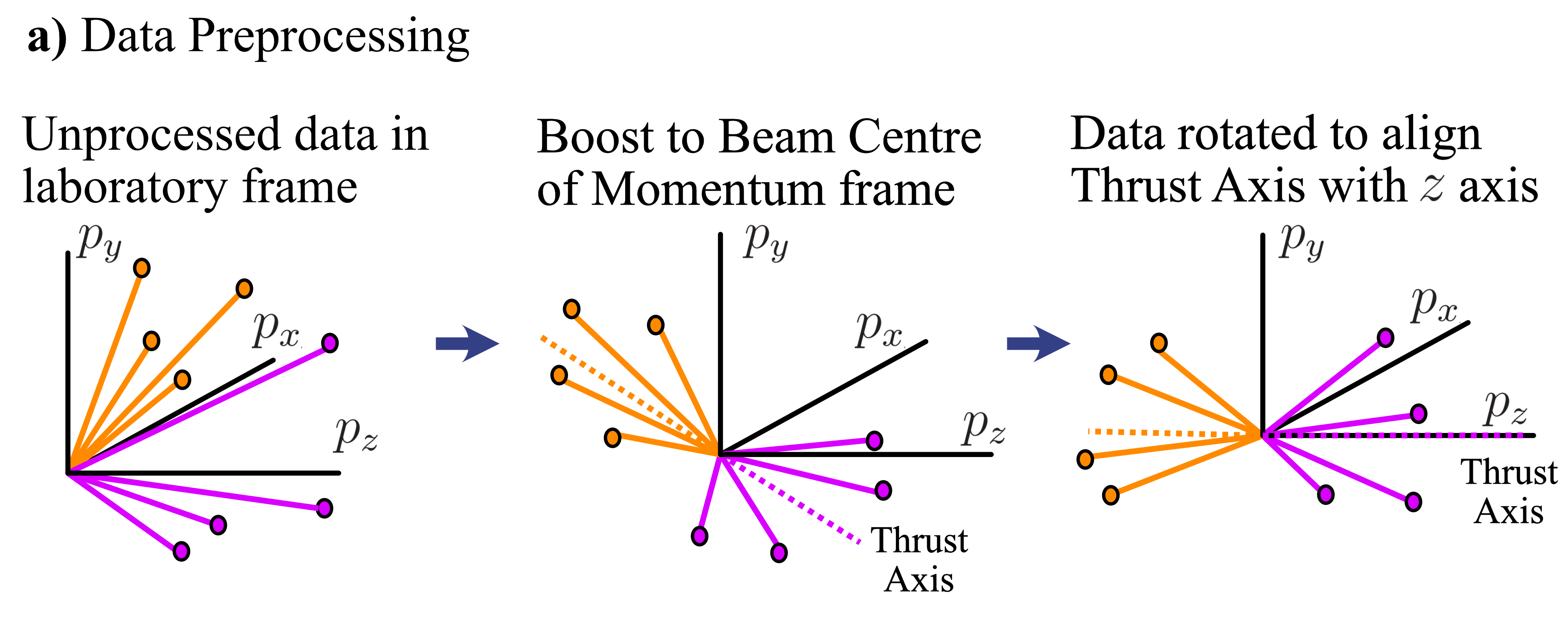}
\caption{An illustration of the preprossessing steps performed on an example $q\bar{q}$ event.}
\label{fig:thrust_explanation}      
\end{figure}

This method focuses on encoding raw momentum data into a quantum state using a quantum circuit. In order to use the raw data in this way, each event is subject to the preprocessing steps illustrated in Figure \ref{fig:thrust_explanation}. We first boost our coordinates to the centre of momentum frame of the event. We then rotate the event such that $\theta = 0$ and $\theta = \pi$ are aligned with the thrust axis of the event. The thrust axis $\mathbf{\hat{n}}$ is calculated by maximizing $T(\mathbf{\hat{n}} )$, defined as
\begin{equation}
    T(\mathbf{\hat{n}}) = \frac{\sum_{i}| \mathbf{P}_i \cdot \mathbf{\hat{n}}|}{\sum_{i}|\mathbf{P}_i|}.
\end{equation}
Where $\mathbf{P}_i$ is a momentum vector belonging to the \textit{i}th particle from either the $B$ candidate or all of the other particles in the event (excluding the $B$ candidate) \cite{PhysRevLett.39.1587}. The momentum of a particle in this frame is represented by three variables in spherical coordinates $(p, \theta, \phi)$ where $p$ is the absolute value of the momentum, $\theta$ is the angle between the particle and the thrust axis and $\phi$ the angle about the thrust axis.

In the centre of mass frame the $e^{+}e^{-} \to q \bar{q}$ pair are formed with substantially more individual momentum than the $e^{+}e^{-} \to \Upsilon(4S) \to B \bar{B}$ pair. Thus the overall distribution of background final state particles are far more ``jet-like'' than the ``spherically'' distributed signal particles. Machine learning algorithms can be trained to distinguish these differences.

Quantum rotation gates accept inputs in the range $[0, 2\pi]$. The $\theta$ and $\phi$ inputs fall naturally into this range. The absolute momentum value $p$ for our data is normalised by a factor $\frac{\pi}{p_{max}}$, where $p_{max}$ is the highest momentum value found in all training and testing datasets, in order for the absolute momentum to take a value between $[0, \pi]$. This ensures the momentum values of $0$ and $p_{max}$ are maximally separated. After preprocessing the data it is then possible to classify the data using a Quantum Support Vector Machine model \cite{havlicek2018supervised}.

\subsection{Quantum Support Vector Machine}
\label{subsec_qsvm}

A classical Support Vector Machine usually works by constructing a hyperplane to separate datapoints that are encoded into a higher dimensional space \cite{Learning, scikit-learn}. In this construction it aims to maximise the distance between the hyperplane and the nearest datapoint of any class. Given $N$ training vectors $\textbf{x}_i$ each with one of two class labels denoted by $y_i \in {1, -1}$, this construction can be achieved by optimising the following

\begin{equation}
    \min_{\alpha}\frac{1}{2}\alpha^T Q \alpha - e^T \alpha; \textrm{subject to }  y^T \alpha =0.
\end{equation}
 Where $e$ is a vector with all elements equal to one. The $\alpha_i$ are referred to as the dual coefficients and are adjusted during the optimisation. $Q$ is an $N$ by $N$ positive semidefinite matrix $Q_{ij} = y_i y_j K(\textbf{x}_i, \textbf{x}_j)$ \cite{sklearn_svm}. The term $K(\textbf{x}_i, \textbf{x}_j) = \phi(\textbf{x}_i)^T \phi(\textbf{x}_j)$ is referred to as the kernel matrix. Each datapoint is transformed into a higher dimension by the encoding $\textbf{x}_j \rightarrow \phi(\textbf{x}_j)$. The explicit form of this encoding is not required by the Support Vector Machine, which only needs to know the kernel matrix that is usually given as an explicit function. 
 
\begin{figure*}
\includegraphics[width=\linewidth,clip]{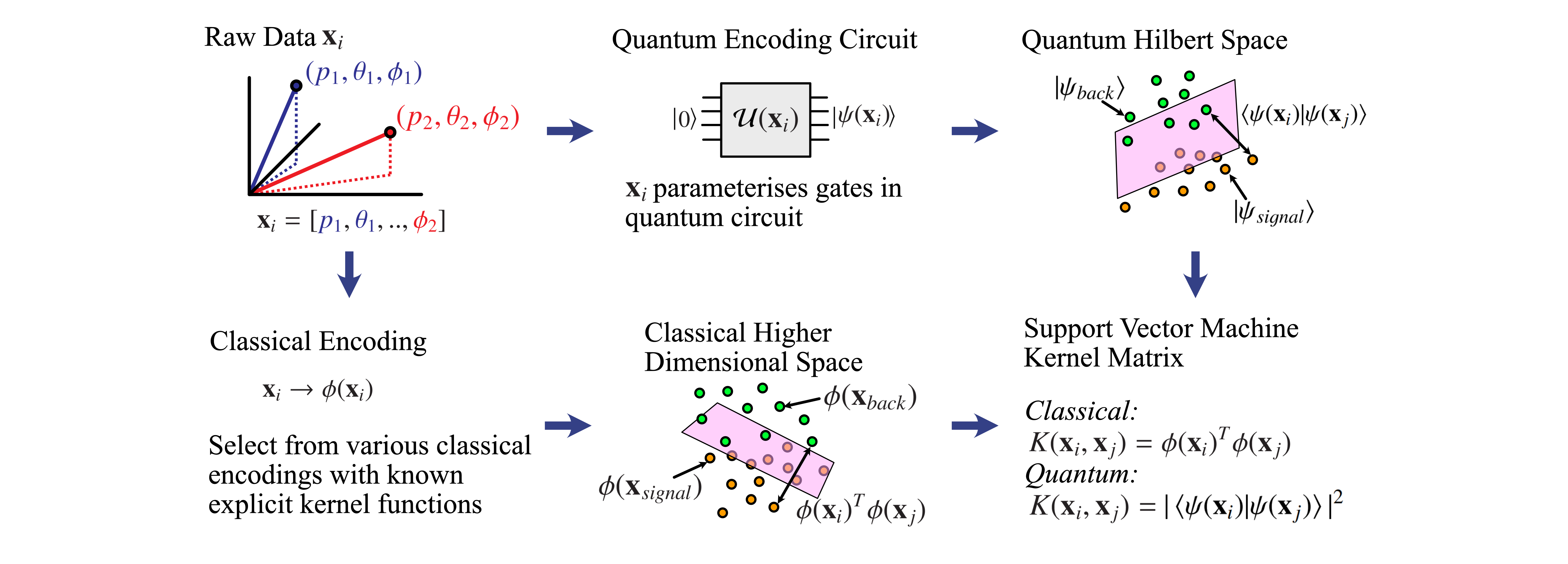}
\caption{(Lower Path) The concept of a classical SVM. Data is encoded in a higher dimensional space where the separation is performed. The inner products between datapoints forms the kernel matrix; this can usually be given as an explicit equation. (Upper Path) The concept of a QSVM. Raw data is encoded into a quantum state using a quantum circuit that is parameterised by the classical data. The separation is made between signal and background in the higher dimensional quantum space. In order to achieve this we need to measure the inner product between the quantum states produced in the circuit.}
\label{fig-overall-picture-explanation}       
\end{figure*}

 For the Quantum Support Vector Machine \cite{havlicek2018supervised} the higher dimensional encoding $\textbf{x}_j \rightarrow \ket{\psi(\textbf{x}_j)}$ is a quantum state which can not be read by a classical algorithm. However, as only the inner product between these quantum encoded states is required it is possible to measure $|\braket{\psi(\textbf{x}_i)|\psi(\textbf{x}_j)}|^2$ which may be used as an estimate for the kernel matrix corresponding to datapoints $\textbf{x}_i$ and $\textbf{x}_j$. The main difference between the quantum and classical support vector machines is that in the classical case an explicit kernel function equation is often known, for example the Radial Basis Function (RBF) kernel defined as $K(\textbf{x}_i, \textbf{x}_j) = \exp \Big(\frac{-|\textbf{x}_i -  \textbf{x}_j|^2}{2 \sigma^2} \Big)$, where $\sigma$ is a free parameter that can be optimised. In the quantum case the kernel matrix is instead calculated using a quantum circuit. The classical and quantum support vector machine approaches are summarised in Figure \ref{fig-overall-picture-explanation}.

The first step of the QSVM is to encode classical data $\textbf{x}_i$ into a quantum state $\ket{\psi(\textbf{x}_i)}$. In Havlicek et al \cite{havlicek2018supervised}, the encoding circuit comprises of gates that are parameterised by the classical data. The aim of this encoding circuit $\mathcal{U}(\textbf{x})$ is to transform each event, a classical $n$ dimensional array consisting of the momentum coordinates of each particle, into a $2^{n}$ dimensional quantum state. The motivation here is that in this higher dimensional Hilbert space it may be easier to separate signal and background events. The input data can be represented as a vector $\textbf{x} \in \mathbb{R}^n$ containing all the momentum variables of the particles considered for an event. This vector is mapped onto a quantum state $\ket{\psi(\textbf{x})}$ using $n$ qubits in the following circuit

\begin{equation}
    \ket{\psi(\textbf{x})} = \mathcal{U}(\textbf{x})\ket{0}^{\otimes n} = \Big( \prod_{i=0}^{L} U(\textbf{x}) H^{\otimes n} \Big)  \ket{0}^{\otimes n}.
\label{encoding_equation}
\end{equation}
An encoding circuit $\mathcal{U}(\textbf{x})$ is applied to an initial $\ket{0}^{\otimes n}$ state. $H^{\otimes n}$ is a parallel implementation of a Hadamard gate on each of the $n$ qubits and $U(\textbf{x})$ is the n-qubit encoding operation. Note that the encoding operation $U(\textbf{x})$ can be repeatedly applied $L$ times to create an $L$ layered encoding circuit, where it is thought that the quantum kernel would be harder to simulate classically as more layers are used. 

\begin{figure}
\includegraphics[width=\linewidth,clip]{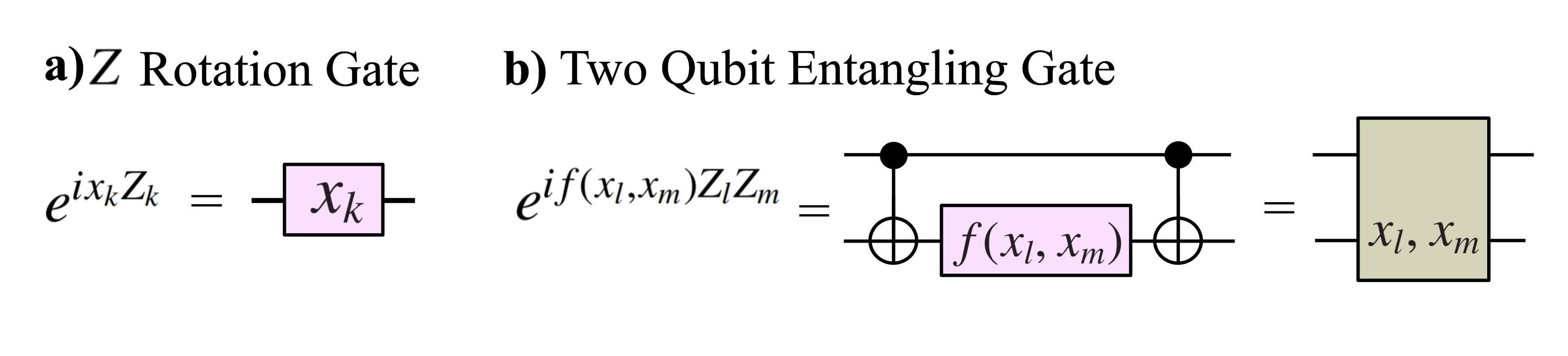}
\caption{ \textbf{a)} The circuit notation for the single qubit part of the encoding gate. The effect of this Z rotation gate is a Z rotation of the qubit's Bloch sphere by an angle $x_k$. \textbf{b)} The circuit notation we use for the two qubit entangling part of the encoding gate. The explicit construction in a real quantum circuit is shown here using Controlled Not gates. This gate effectively introduces some form of phase shift between two qubits that depends on some function of the respective momentum variables in the classical data $x_l$ and $x_m$.}
\label{fig-entangling-specification}       
\end{figure}
\begin{figure}
\includegraphics[width=\linewidth,clip]{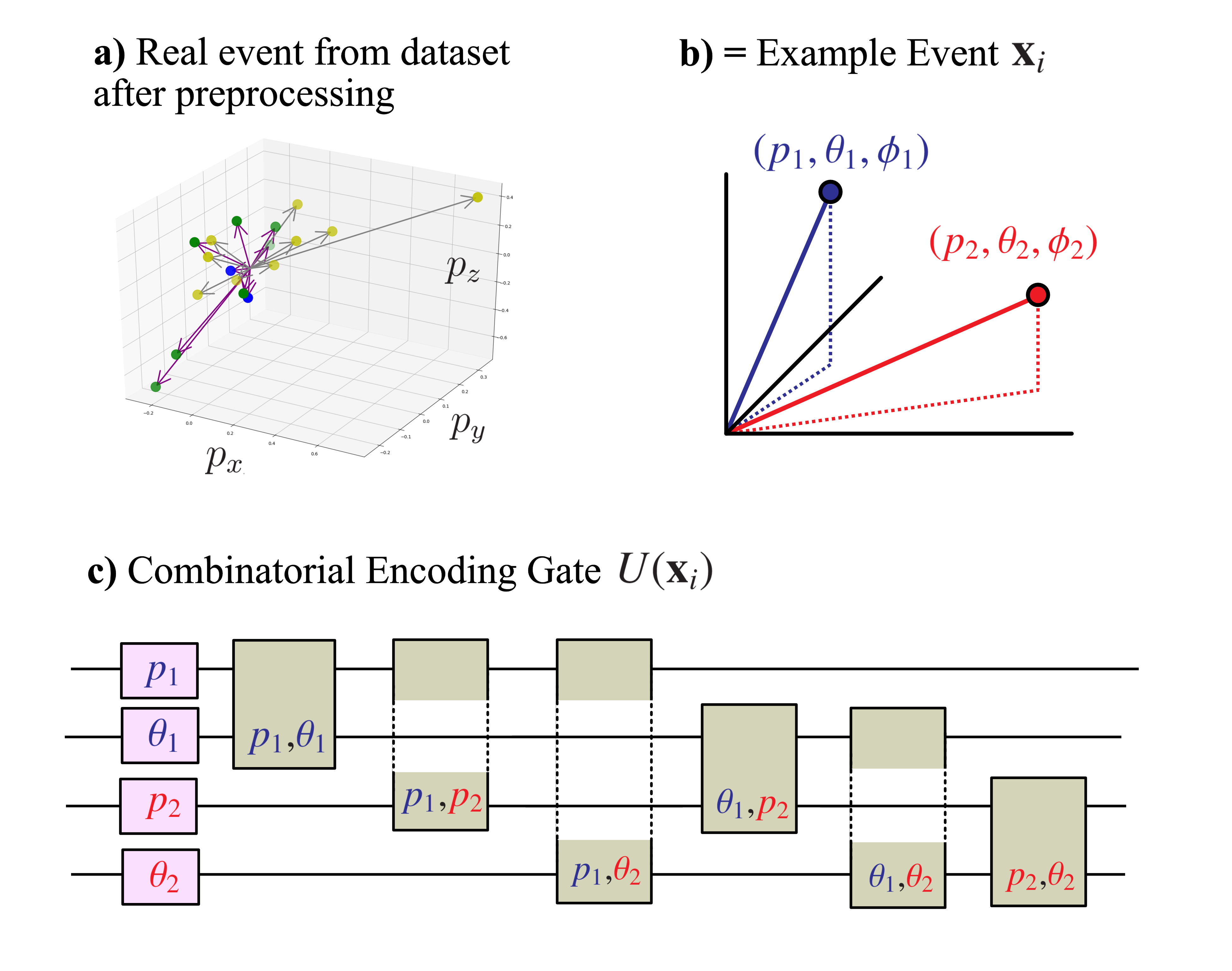}
\caption{\textbf{a)} An event from the real dataset after preprocessing is shown. Different coloured points correspond to different types of charged particles. \textbf{b)} An example particle decay event containing two particles demonstrating the raw momentum data in spherical coordinates. \textbf{c)} The circuit structure for the combinatorial encoding operation $U(\textbf{x})$ \cite{havlicek2018supervised}. For readability the circuit omits the $\phi$ variable.}
\label{fig-qsvm-new}       
\end{figure}

\subsection{Combinatorial Encoding}

The encoding operation $U(\textbf{x})$ that was used in the technique suggested by Havlicek et al \cite{havlicek2018supervised} is illustrated in Figure \ref{fig-qsvm-new}. It is formally defined as
\begin{equation}
    U(\textbf{x}) = \exp \Big( i \sum_{k=1}^n x_k Z_k + i \sum_{l = 1}^{n - 1} \sum_{m > l}^n f(x_l, x_m) Z_l Z_m   \Big),
\end{equation}
where $Z_k$ represents the Pauli Z matrix applied to qubit $k$. This circuit first introduces a phase shift to each individual qubit by an amount $x_k$. This is followed by an entangling step where the function $f$ effectively quantifies some form of a phase shift between the two qubits that are to be entangled. The entangling function we used is
\begin{equation}
    f(x_l, x_m) = \frac{1}{\pi}(x_l - \pi)(x_m - \pi).
\end{equation}
This circuit explicitly entangles every qubit with every other qubit, meaning all combinations of qubits are entangled. We will therefore refer to this circuit throughout this paper as the combinatorial encoding circuit \cite{havlicek2018supervised}.

The purpose of the encoding operation is to turn classical data into a quantum state. In order to use a support vector machine we need to calculate the inner product between every event in this quantum space $\braket{\psi(\textbf{x}_i)|\psi(\textbf{x}_j)}$. By referring to equation (\ref{encoding_equation}) this quantity can be written as
\begin{multline}
\braket{\psi(\textbf{x}_i)|\psi(\textbf{x}_j)} = \\ \bra{0} H^{\otimes n} U^{\dag}(\textbf{x}_i) H^{\otimes n} U^{\dag}(\textbf{x}_i) U(\textbf{x}_j) H^{\otimes n} U(\textbf{x}_j) H^{\otimes n} \ket{0}.  
\end{multline}

\begin{figure}
\includegraphics[width=\linewidth,clip]{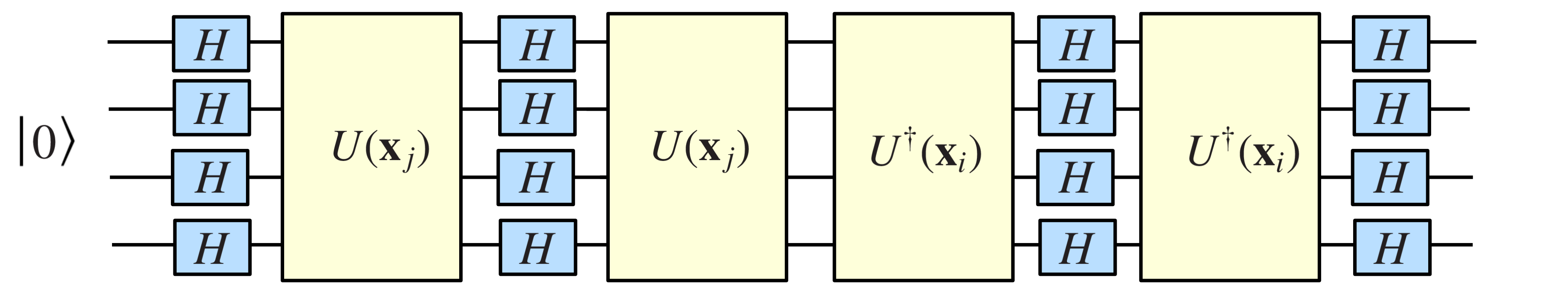}
\caption{The general circuit for estimating the inner product of two quantum encoded states. The probability of measuring $\ket{0}^n$ as the final state of this circuit will estimate the quantity $|\braket{\psi(\textbf{x}_i)|\psi(\textbf{x}_j)}|^2$.}
\label{fig:kernel_estimate_circuit_full}       
\end{figure}

This kernel estimation circuit, which is illustrated in Figure \ref{fig:kernel_estimate_circuit_full}, acts to determine the inner product between the two quantum states. The circuit is run repeatedly over many shots (identical runs) and the proportion of $\ket{0}^n$ state measurements is calculated. The proportion of $\ket{0}^n$ measurements is an estimate for the probability $|\bra{0} H^{\otimes n} U^{\dag}(\textbf{x}_i) H^{\otimes n} U^{\dag}(\textbf{x}_i) U(\textbf{x}_j) H^{\otimes n} U(\textbf{x}_j) H^{\otimes n} \ket{0}|^2. \phantom{A}$ This process therefore produces an estimate for the quantity $|\braket{\psi(\textbf{x}_i)|\psi(\textbf{x}_j)}|^2$ which can then be used as the kernel matrix entry for two events $\textbf{x}_i$ and $\textbf{x}_j$. This is repeated for all combinations of events in the dataset until a full kernel matrix is obtained. This kernel is then passed to a classical support vector machine to perform the classification. The overall effect of this is to take events of dimension $n$ and project them into a $2^n$ dimensional quantum space where the separation is then performed.

\section{Alternative Encoding Methods}

The combinatorial encoding circuit that is shown in Figure \ref{fig-qsvm-new} corresponds to a specific kernel. By making adjustments to this encoding circuit we are able to construct entirely different kernels, which have the potential to perform better on our dataset while also using fewer gates and qubits. 

\subsection{Bloch Sphere Encoding}

\begin{figure}
\includegraphics[width=\linewidth]{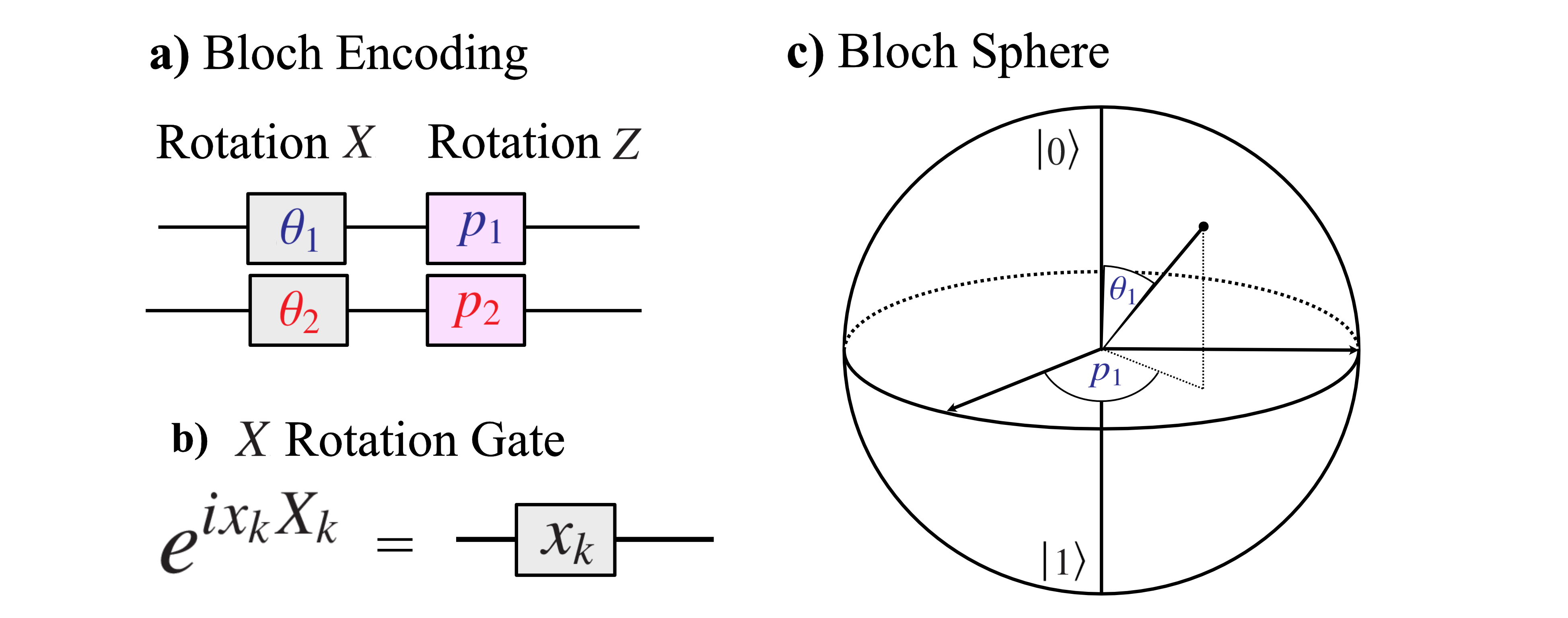}
\caption{ \textbf{a)} An encoding operation that encodes two variables of a particle into a single qubit. \textbf{b)} The definition of the X rotation gate. $X_k$ is the Pauli $X$ matrix applied to qubit $k$. \textbf{c)} The resulting Bloch sphere of a qubit, if initially in the $\ket{0}$ state, after being acted on by this circuit.}
\label{fig:bloch_sphere}
\end{figure}

In an effort to encode the same amount of classical information into fewer qubits we implemented a model that encodes the $\theta$ and $p$ variables of each particle into a single qubit. There are several ways in which multiple classical variables can be encoded into a single qubit \cite{LaRose_2020}. To test this concept we construct a circuit that applies an $X$ rotation by $\theta$, followed by a $Z$ rotation by $p$ to the Bloch sphere of the qubit. This encodes the two variables into the Bloch sphere of a single qubit as shown in Figure \ref{fig:bloch_sphere}. Note that this circuit contains no quantum entanglement.

\subsection{Separate Particle Encoding}

\begin{figure}[h!]
\includegraphics[width=\linewidth]{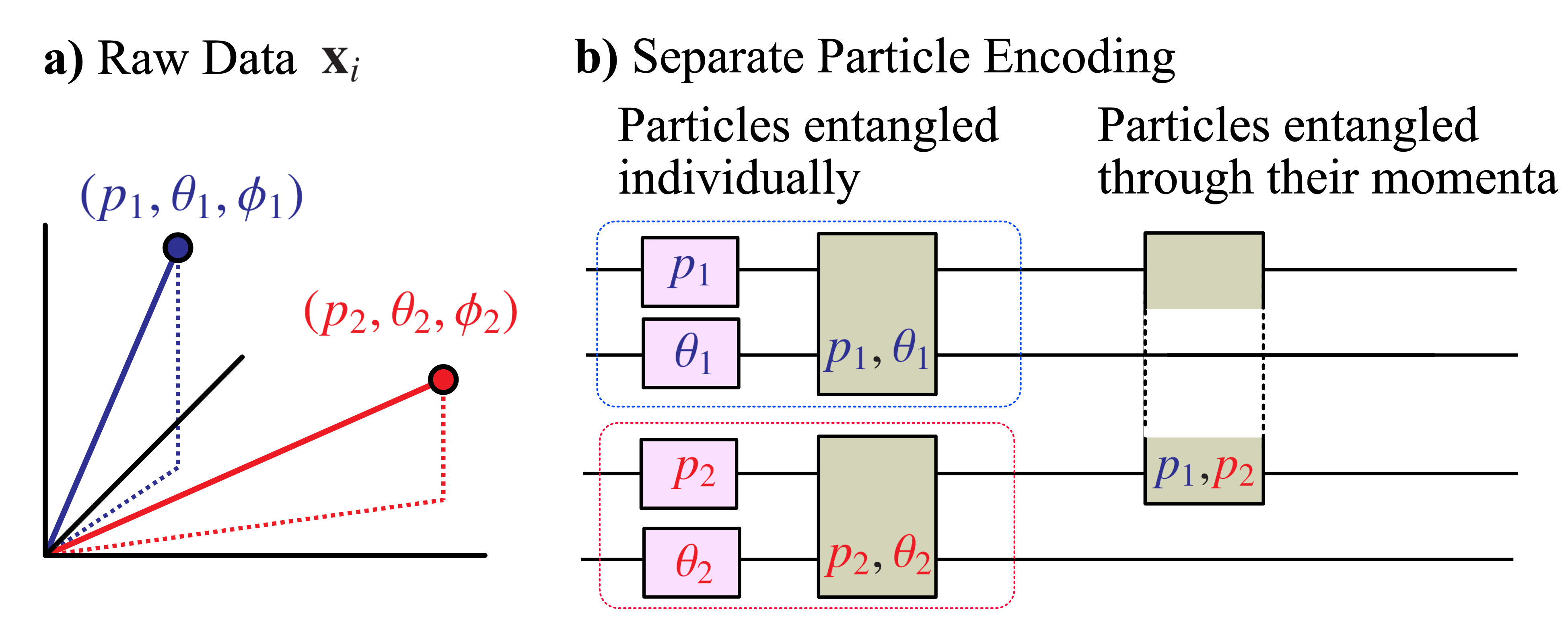}
\caption{ \textbf{a)} The raw data classical input for a two particle event. \textbf{b)} The separate particle encoding operation $U(\textbf{x})$. Initially a 2 qubit entangled state is produced for each particle based on each particle's momentum variables. After individual particles are separately encoded, each particle is then entangled with every other particle through one of its qubits.}
\label{fig:particle_encoding}
\end{figure}

The structure of the encoding circuit directly affects the kernel function and the final classification result. One possible issue with the combinatorial encoding circuit is that it treats every classical variable identically; $p_1$ would be entangled with $\theta_1$ in the same manner as $p_1$ is entangled with $p_2$, despite $p_2$ being a variable from a different particle. The circuit has no built-in way of discriminating the individual particles. Considering this we introduce a separate particle encoding circuit. The first layer involves entangling the momentum variables for each particle individually, resulting in a 2 qubit quantum state for each particle. This is followed by a layer that entangles the states representing each particle with every other particle based on their momenta. This separate particle encoding operation $U(\mathbf{x})$ is shown in Figure \ref{fig:particle_encoding} for the 2 particle case.

This technique has the advantage of using fewer quantum gates in total than the combinatorial circuit, which reduces the error rate when running on a quantum device. It could also be adapted to include additional information about each particle, by expanding the number of qubits to include variables such as charge, mass etc. without as significant an increase in the number of gates used compared to the combinatorial encoding circuit. 

\subsection{Separate Particle with Bloch Encoding}

We can merge the Bloch sphere encoding idea into the separate particle circuit, by encoding both angles $\theta$ and $\phi$ into the same qubit. This results in all three momentum variables for a particle being encoded into 2 qubits. In this encoding gate the $\theta$ and $\phi$ angles of each particle are encoded into the $\theta$ and $\phi$ angle rotations of one qubit, and the momentum of the particle is encoded into the other qubit. Entanglement between these 2 qubits can optionally be introduced here, for our simulations with noise and real machine run we used a two qubit entangling gate as shown in Figure \ref{fig-entangling-specification} where the phase shift was given by $f(p, \theta, \phi) = \frac{1}{\pi^2}(\pi - p)(\pi - \theta)(\pi - \phi)$. In the next step each particle, now represented by 2 qubits, is entangled with every other particle via their momentum qubit in the same way as the separate particle encoding operation in Figure \ref{fig:particle_encoding}.

\section{Result Comparisons}
\label{section:result_compar}

We have limited our simulations to events that contain exactly 4 particles, which accounts for around $14\%$ of the total events. We could generalise this approach by building circuits for different numbers of particles and then combining the end results. However, for the purposes of this exploratory analysis we will only consider 4 particle events. The simulations were run with 60,000 training events and 10,000 testing events using the Qiskit \textit{statevector\_simulator} in the absence of noise \cite{Qiskit}. This was repeated 10 times in total using a random selection of events for the training and testing datasets each time. The results of the different encoding circuits are summarised in Table \ref{tab-full}. 

\begin{table}
\caption{Average results from 10 random dataset samples obtained by classically simulating various encoding circuits using Qiskit \textit{statevector\_simulator} with 60,000 training events and 10,000 testing events in each sample. The uncertainty on each of the mean values stated is $\pm$ 0.001.}
\label{tab-full}
\begin{tabular}{lll}
\hline\noalign{\smallskip}
Encoding Circuit & Accuracy & AUC \\
\noalign{\smallskip}\hline\noalign{\smallskip}
Combinatorial Encoding & 0.762 & 0.822 \\
Separate Particle Encoding & 0.776 & 0.835\\
Bloch Sphere Encoding & 0.764 & 0.836\\
Separate Particle with Bloch & 0.771& 0.848\\
\noalign{\smallskip}\hline\noalign{\smallskip}
Classical RBF Kernel SVM & 0.728& 0.793\\
XGBoost & 0.590 & 0.621 \\
\noalign{\smallskip}\hline
\end{tabular}
\end{table}

These results suggest that the physically motivated improvements made in the separate particle circuit have a positive impact on the discriminatory power of the circuit compared to the combinatorial encoding. Furthermore, combining this method with the Bloch sphere encoding resulted in our best average AUC score of 0.848. For comparison we tested classical SVMs on the data for various different classical kernels with the best result being the radial basis function (RBF) Kernel with an average AUC of 0.793. We also included XGBoost as a comparison, deciding on the hyperparameters (maximum tree depth, number of estimator, minimum split loss and learning rate) using GridSearchCV \cite{scikit-learn}. The uncertainty in the mean values was $0.001$ for each entry in Table \ref{tab-full}, this uncertainty was calculated by taking the standard deviation and dividing by the square root of the number of samples (which is 10 in this case). This suggests the dataset size used in these simulations was large enough to enable statistically significant comparisons between encoding circuits.

\section{Simulations with Noise and Error Mitigation}
\label{section:noise_sim}

The simulation results demonstrate the potential for quantum algorithms on ideal noiseless devices. Modern quantum devices do however have noticeable error rates which are dependent on factors such as the number of gates used in the circuit. To investigate the effect this has on our results we re-run our simulation with a noise model \cite{Qiskit}. In order to reduce the impact of these errors we implement a measurement error mitigation technique \cite{error_mitigation} recently demonstrated in the context of verifying whole-array entanglement in the IBM Quantum \textit{ibmq\_manhattan} device \cite{mooney2021wholedevice}. This procedure consists of performing an initial calibration of the basis states for the noisy device. The noisy basis state measurements are used to construct a matrix; the inverse of this matrix when applied to a noisy basis state should take it to the ideal basis state. By applying this calibration matrix to our final results we can reconstruct a final measurement closer to the ideal noiseless case.

The separate particle with Bloch encoding circuit was tested using events that contained 3 particles only, which accounts for $20\%$ of total events. As mentioned previously, this could be generalised by building circuits with more qubits for events with more particles and combining the results. However, for this section we focused on events that contained only 3 particles so that the circuits could be run on a 6 qubit quantum machine. The Qiskit \textit{qasm\_ simulator} is used with a simulated noise model based on the IBM Quantum {\textit{ibmq\_toronto}} device. The classification was repeated with 10 different datasets, each consisting of 30 training and 30 testing points to find an average classification performance. The separate particle with Bloch encoding circuit achieved an average accuracy of 0.670 $\pm$ 0.100 and an average AUC of 0.751 $\pm$ 0.100. For comparison, this test repeated with an ideal noiseless simulation gives an accuracy of 0.750 $\pm$ 0.050 and an AUC of 0.789 $\pm$ 0.110. 

The average AUC score for the separate particle with Bloch encoding circuit is 0.751 for this dataset size when run with simulated noise. In contrast, repeating this investigation for the combinatorial encoding circuit results in an average accuracy of 0.530 $\pm$ 0.086 and an average AUC of 0.550 $\pm$ 0.131, a significant reduction in performance. This could possibly be due to the greater number of gates used in the combinatorial encoding circuit compared to the separate particle with Bloch encoding circuit, resulting in a larger source of quantum error.

\section{Experimental Testing on Real Quantum Devices and Comparison}

 The separate particle with Bloch encoding circuit was tested on a real device using 3 particle events only. The dataset consisted of 30 training points and 30 testing points. This was run using 6 qubits on the IBM Quantum {\it ibmq\_casablanca} device and repeated 10 times for the same dataset achieving an average accuracy of 0.640 $\pm$ 0.036 and an average AUC of 0.703 $\pm$ 0.063, where the quoted uncertainties in this case use the standard deviation of multiple runs using the same dataset to demonstrate the effect of quantum noise in the real machine.
 
The upper section of Table \ref{tab-real} shows the comparison of ideal simulations of varying training dataset sizes alongside a simulated noise model of the IBM Quantum {\it ibmq\_ toronto} device. These results are averaged over 10 random datasets. The uncertainties quoted are the standard deviation of the distribution of trials. The result for the real device was a single run and is consistent within uncertainty of the simulated noise runs. 

\begin{table}
\caption{Results obtained using the separate particle with Bloch encoding circuit as the training dataset size is varied. The testing dataset size was fixed at 30. Simulations were run using Qiskit then repeated and averaged over 10 different datasets. The Simulated Noise result includes a classical simulation of the noise found in the IBM Quantum {\it ibmq\_toronto} device. The Real Device result refers to a single dataset being run on the IBM Quantum {\it ibmq\_casablanca} device.}
\label{tab-real}
\begin{tabular}{llll}
\hline\noalign{\smallskip}
Device Type & Training & Accuracy & AUC \\
\noalign{\smallskip}\hline\noalign{\smallskip}
Ideal Simulation & 1000 & 0.77 $\pm$ 0.03 & 0.83 $\pm$ 0.05 \\
Ideal Simulation & 100 & 0.75 $\pm$ 0.05 & 0.78 $\pm$ 0.04 \\
Ideal Simulation & 30 & 0.75 $\pm$ 0.05  & 0.79 $\pm$ 0.11\\
Simulated Noise & 30 & 0.67 $\pm$ 0.10 & 0.75 $\pm$ 0.10 \\
\noalign{\smallskip}\hline\noalign{\smallskip}
Real Device & 30 & 0.64 & 0.70 \\
\noalign{\smallskip}\hline
\end{tabular}
\end{table}

The trend demonstrates that smaller training datasets result in worse AUC scores with larger uncertainties. These results suggest that for small dataset sizes there will be rather large uncertainties, even in the absence of quantum noise, due to the size of the training dataset itself. As availability of quantum machines increases it is hoped that larger datasets could be run, in which case the main source of uncertainty would be expected to derive from quantum noise instead.


\section{Conclusion}

We have demonstrated on a small scale how quantum machine learning may be applied to signal-background classification for continuum suppression in the study of B mesons. Simulating the combinatorial encoding circuit on the signal-background classification problem we measured an average AUC of 0.822, outperforming the classical SVM and XGBoost methods tested for this dataset. The separate particle with Bloch encoding circuit, which we designed for particle data, improved the average AUC further to 0.848. This encoding method also used fewer qubits and quantum gates than the combinatorial circuit.
Using a smaller dataset in the presence of simulated quantum noise the separate particle with Bloch encoding method achieved an average AUC of 0.750. In contrast, the combinatorial circuit in the same set-up performed significantly worse in the presence of simulated noise with an AUC of 0.550, a possible explanation of this could be the much higher number of gates in the combinatorial circuit compared to the separate particle Bloch encoding circuit. Running the separate particle encoding circuit on a real quantum device resulted in an AUC score of 0.703, which lies within the error range of the simulated noise runs. Further work could involve using quantum error mitigation techniques \cite{PhysRevX.8.031027} to improve the performance on real quantum devices in the presence of noise.

When using a limited number of inputs the QSVM outperformed classical methods in simulations. This result suggests that the quantum kernel created by the circuit is useful for signal background classification for certain datasets. We demonstrate a separate particle Bloch encoding circuit that performed best in our noiseless simulations and performed significantly better than the combinatorial circuit in the presence of simulated noise. The number of input data sets and performance in the quantum approach was limited due to the size and error rates of current quantum computers. If expanded to use more inputs and larger dataset sizes it is foreseeable that the QSVM may be able to compete with current state-of-the-art classical techniques, which can achieve AUCs of 0.930 \cite{HAWTHORNEGONZALVEZ201954}. There are a plethora of alternative quantum machine learning methods that could utilise the encoding circuits discussed here. There may also be other data classification tasks, from particle physics or elsewhere, that could benefit from this method of encoding. Whether the same encoding circuits that performed well for the QSVM technique also succeed in other approaches would be a question for further investigation.
\\

\renewcommand{\abstractname}{Acknowledgements}
\begin{abstract}
 This work was supported by the University of Melbourne through the establishment of an IBM Quantum Network Hub at the University. This research was supported by the Australian Research Council from grants DP180102629 and DP210102831. JH acknowledges the support of the Research Training Program Scholarship and the N.D. Goldsworthy Scholarship. CDH is supported by a research grant from the Laby Foundation.
\end{abstract}

\renewcommand{\abstractname}{Conflict of interest}
\begin{abstract}
 On behalf of all authors, the corresponding author states that there is no conflict of interest.
\end{abstract}

\renewcommand{\abstractname}{Data Availability}
\begin{abstract}
 The data that supports the findings of this study is from the Belle II experiment but restrictions apply to the availability of this data and so it is not publicly available. Data is available from the authors upon reasonable request and with permission from the Belle II experiment.
\end{abstract}

%
%

\bibliographystyle{spmpsci}      
\bibliography{references}   

%
%

\end{document}